\documentclass{vgtc}                          % final (conference style)
\graphicspath{{figures/}{pictures/}{images/}{./}} % where to search for the images

\usepackage{times}                     % we use Times as the main font
\usepackage{calc}
\usepackage{enumitem}
\usepackage{csquotes}% to modify bullet lists

\usepackage{mathptmx}                  % use matching math font

\onlineid{0}

\vgtccategory{Research}

\vgtcinsertpkg

\newif\ifshowchanges\showchangesfalse
\newcommand{\inserted}[1]{%
	\ifshowchanges
    	\textcolor{red}{{}#1}%
	\else
    	#1%
	\fi}
\newcommand{\deleted}[1]{%
	\ifshowchanges
    	\textcolor{white!60!black}{{}#1}%
	\else\fi}

\title{EventColumn: Integrating Event Sequences into Tabular Visualizations}

\author{Jakob Zethofer\thanks{E-mail: jakob.zethofer@pro2future.at}\\%
     \scriptsize Pro2Future GmbH%
\and Andreas Hinterreiter\thanks{E-mail: andreas.hinterreiter@jku.at}\\%
     \scriptsize JKU Linz%
\and Lukas Schiefermüller\thanks{E-mail: lukas.schiefermueller@voestalpine.com}\\%
    \scriptsize voestalpine Stahl GmbH
\and Belgin Mutlu\thanks{E-Mail: belgin.mutlu@pro2future.at}\\%
    \scriptsize Pro2Future GmbH%
\and Marc Streit\thanks{E-mail: marc.streit@jku.at}\\%
     \scriptsize JKU Linz}

\teaser{
  \centering
  \includegraphics[width=\linewidth]{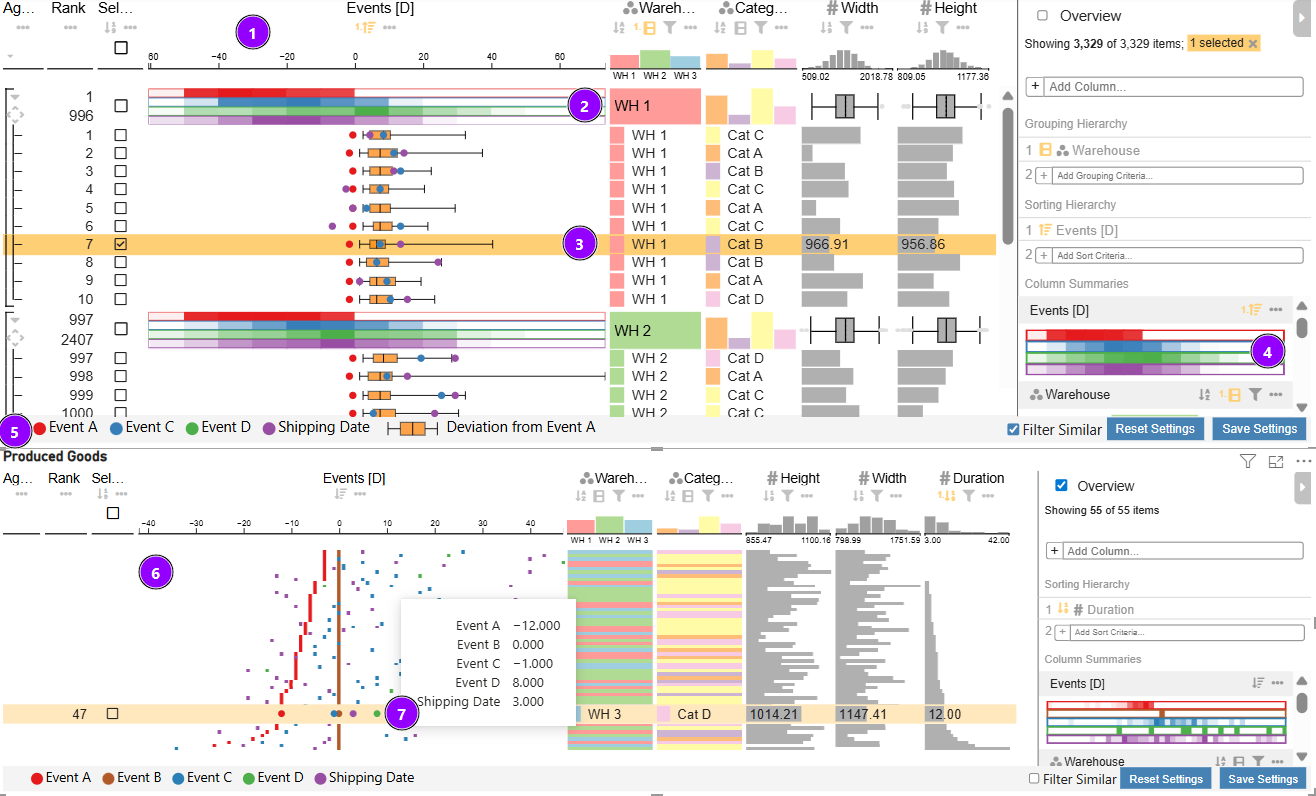}
  \caption{EventColumn provides visual decision support for event-based data, for example, in production logistics, where it helps users estimate the storage duration of production items. 
  EventColumn displays event sequences alongside conventional tabular attributes (e.g., warehouse location) in a single table row.
  Events are displayed on a shared timeline~(1). Grouping event rows by type summarizes them in heatmaps~(2). A boxplot shows the distribution of specific event stages~(3), in this case the storage duration of already produced items relative to the current production event. The column summary~(4) shows the distribution of all events.  EventColumn allows comparison of multiple event sequences by aligning and sorting them by event type (5). When users select an item in the main table, a linked table~(6) is filtered to show details for items represented by the boxplots. Hovering over an item reveals detailed event timestamps in a tooltip~(7).}
  \label{fig:teaser}
}

\abstract{%
We introduce EventColumn, a new column type that integrates event-sequence data with heterogeneous tabular attributes into a single unified table. EventColumn lets analysts compare event sequences alongside numerical, categorical, and temporal attributes at both instance and group levels, offering a compressed overview, heatmap group summaries, alignment by event types, and boxplots of similar historical items. We developed EventColumn together with collaborators from the steel industry to facilitate the analysis of production events and warehouse logistics, but the solution \deleted{generalizes to a wide range of event sequence datasets with additional tabular attributes}\inserted{applies to other event-sequence domains}. Unlike most existing approaches that compare either event sequences or tables, EventColumn supports simultaneous comparison of both. We demonstrate its integration with Taggle and Microsoft Power BI on data from steel production logistics and on a public e‑commerce dataset.%
}


\begin{document}


\firstsection{Introduction}

\maketitle
\label{sec:introduction}
In multi-stage manufacturing settings, successive production stages are often distributed across different locations and points in time, with products stored in warehouses between stages.
When a warehouse reaches capacity, items must be relocated to external storage facilities.
Optimal relocation decisions require identifying items that will be stored the longest to reduce the total number of transports and avoid damage from additional handling.
Such relocation decisions can benefit from a visual decision support system.

We developed our approach in collaboration with a steel manufacturer, focusing on logistics between the hot rolling and pickling (i.e., coil surface acid treatment) production steps.
In the current process, logistics employees assess which coils to move to outdoor storage by evaluating the remaining storage duration based on the coil's characteristics and relevant events, such as the shipping date and the hot rolling date.
The expected storage duration before pickling is often unclear, as pickling is typically scheduled only about three days in advance, and coils require the same cooling period after hot rolling.
Training a machine learning model for predicting storage duration did not yield reliable results\deleted{, as }\inserted{. We experimented with multiple different clustering algorithms, regression and classification models, and inspected the most promising results with a detailed SHAP analysis~\cite{lundberg_unified_2017}. The results suggested that }human factors and non-standardized planning decisions \deleted{are}\inserted{were} not fully captured in historical data.
We therefore designed a visual analytics approach that keeps the human in the loop by providing domain experts with a dashboard that helps them estimate a coil's storage duration.

Supporting this task requires comparing candidate coils across two types of data: (\textit{i})~\emph{event sequences} containing the dates of production-related events such as hot rolling completion, planned galvanizing, or shipping, and (\textit{ii})~\emph{heterogeneous tabular attributes} such as steel category, coil dimensions, and warehouse location.
Comparing each candidate against similar historical coils whose actual storage duration is known further aids decision-making.
While event sequence visualization~\cite{guo_survey_2022} and tabular visualization~\cite{dimara_conceptual_2018} are both well-researched areas, existing tools do not display event sequences and heterogeneous tabular attributes together in a single table. Instead, they separate them into different views or tabs~\cite{van_dortmont_chronocorrelator_2019,fails_visual_2006}, which makes comparison of items more difficult.

We present \textbf{EventColumn}, a new column type that displays event sequences alongside numerical, categorical, and temporal attributes.
% Events are rendered as colored circles per row on a shared, interactive timeline. A distribution boxplot overlaid in the same column summarizes the distribution of the target quantity across similar historical items.
% The EventColumn offers an overview mode that reduces row height and displays heatmap aggregations for each event type. Events can be aligned and sorted by type, with tooltips displaying values on hover.
To demonstrate our approach, we integrate EventColumn with Taggle~\cite{furmanova_taggle_2020}, a tabular visualization technique, and Microsoft Power BI~\cite{microsoft_powerbi_2026}. 
We showcase the functionality of EventColumn on a steel production logistics dataset and demonstrate its generalizability on an e-commerce dataset.
% We demonstrate the EventColumn in a decision support dashboard for steel production logistics and show its applicability to a publicly available e-commerce dataset.

\section{Related Work}

We position our work at the intersection of visual analytics for decision support in manufacturing, event sequence visualization, and tabular data visualization.

\subsection{Visual Analytics for Decision Support}

Visualization plays an increasingly important role in industrial decision-making.
Zhou et al.~\cite{zhou_survey_2019} surveyed visualization approaches for smart manufacturing, identifying the production phase---where our work is situated---as one of the key stages that benefits from visual analytics.
Schreck et al.~\cite{schreck_visual_2023} discussed challenges of deploying visual data science in industrial settings, highlighting the need to cope with the four Vs of big data (volume, velocity, variety, veracity) and to integrate domain expertise.
Tools like ViDX~\cite{xu_vidx_2017} and LiveGantt~\cite{jo_livegantt_2014} address assembly-line diagnostics and scheduling visualization, respectively, but focus on the overall production process rather than inter-step logistics.

\subsection{Event Sequence Visualization}

Event sequence visualization has been extensively surveyed.
Guo et al.~\cite{guo_survey_2022} provide a comprehensive design space covering data scales, visual representations, and interaction techniques.
Van der Linden et al.~\cite{van_der_linden_survey_2023} extend this design space by including dimensions such as granularity, positioning, and alignment for comparative analysis.
Yeshchenko and Mendling~\cite{yeshchenko_survey_2024} propose the ESeVis framework, distinguishing between instance and model representations across different visualization types.

Existing tools include LifeLines~2~\cite{wang_aligning_2008}, which aligns and ranks event records by sentinel events;
EventFlow~\cite{monroe_temporal_2013}, which aggregates and simplifies temporal event sequences;
and Similan~\cite{wongsuphasawat_finding_2009}, which supports similarity-based comparison of categorical event sequences.
Several systems augment events with additional data:
ChronoCorrelator~\cite{van_dortmont_chronocorrelator_2019} enriches events with global density plots and filter widgets, and EmoCo~\cite{zeng_emoco_2020} combines event timelines with bar code charts summarizing emotions.
However, existing approaches separate event visualizations from tabular attribute displays: for example,
TempViz~\cite{fails_visual_2006} uses separate tabs for events and attributes,
while Guo et al.~\cite{guo_comparative_2020} show events and multivariate data in separate linked views.
None of these tools unify event sequences and heterogeneous tabular columns within a single table layout.

\subsection{Tabular Visualization}

Dimara et al.~\cite{dimara_conceptual_2018} found that tabular visualizations slightly outperformed parallel coordinates and scatterplots for decision-support tasks involving multi-attribute comparison. Bartram et al.~\cite{bartram_tables_2022} highlighted the importance of tables to many users for viewing the unaggregated data. Both of these studies motivate our choice of a table-based layout.
LineUp~\cite{gratzl_lineup_2013} introduced interactive visual analysis of multi-attribute rankings, encoding numerical values as bar charts within table columns and supporting flexible sorting and weighting.
Taggle~\cite{furmanova_taggle_2020} extends LineUp with grouping, aggregation, and an overview mode that compresses rows to a single pixel height, enabling the display of hundreds of items simultaneously.
Our work extends Taggle by introducing a new EventColumn that embeds event-sequence data directly into the table, thereby combining the strengths of tabular and event-sequence visualization in a unified layout.

\section{Task Abstraction and Design Goals}
\label{sec:task-abstraction-and-design-goals}

In this section, we characterize our collaborators' existing workflow, derive abstract analysis tasks, and discuss the resulting design goals for EventColumn.

\subsection{Application Context}

We gathered the details of the decision-making process for coil relocation described in Section~\ref{sec:introduction} through multiple meetings and a plant tour with employees from the logistics department.
Based on our analysis, we identified a four-step target workflow.

In \textbf{Step~1}, the logistics employee focuses on coils stored in a warehouse approaching capacity, excluding those not suitable for relocation (e.g., due to urgency flags).
\textbf{Step~2} involves identifying coils likely to require a long storage period, based on criteria such as a shipping date far in the future.
In \textbf{Step~3}, the remaining candidate coils are compared in more depth: categorical, numerical, and temporal attributes are evaluated alongside the distribution of storage durations of similar coils from historical data.
%TODO: remove step 4?
In \textbf{Step~4}, selected coils are scheduled for transport to the external storage location.

\subsection{Task Abstraction}

We performed a task abstraction~\cite{munzner_visualization_2014} of the identified workflow, resulting in the following abstract visualization tasks:

\begin{enumerate}[
    label=\textbf{T\arabic{*}},
    leftmargin=0pt,
    labelwidth=\widthof{\textbf{T2}},
    itemindent=\widthof{\textbf{T2}}+0.5em,
    labelsep=0.5em,
    itemsep=.5ex,
    ]
    \item\label{task:filter-candidates} \textbf{Filter items based on user-defined characteristics.} For example, the logistics employee browses the list of current coils and identifies candidates for relocation based on known criteria, such as a shipping date far in the future or a particular product type.
    \item\label{task:select-compare} \textbf{Select items through comparison} with items from the present and past via four subtasks:
    \begin{enumerate}[
        label*=\textbf{.\alph{*}},
        leftmargin=1em,
        labelwidth=\widthof{\textbf{T2.a}},
        itemindent=\widthof{\textbf{T2.a}}+0.5em,
        labelsep=0.5em,
        itemsep=.5ex,
        ]
        \item\label{task:derive-similar} \textbf{\inserted{Inspect}\deleted{Derive} similar items} from historical data based on user-defined attributes.
        \item\label{task:summarize} \textbf{Summarize the target quantity} (e.g., storage duration) of \deleted{derived }similar items.
        \item\label{task:compare-attributes-similar} \textbf{Compare attributes of similar items} (categorical, numerical, and temporal) to assess the reliability of the estimate.
        \item\label{task:compare-sttributes-candidate} \textbf{Compare attributes of candidate items} to select the most promising ones (discover--explore--compare). This includes event sequences, target quantity distribution, and other attributes.
    \end{enumerate}
\end{enumerate}

\subsection{Design Goals}

We identified the following design goals based on the defined abstract tasks and target workflow.

\begin{enumerate}[
    label=\textbf{D\arabic{*}},
    leftmargin=0pt,
    labelwidth=\widthof{\textbf{D2}},
    itemindent=\widthof{\textbf{D2}}+0.5em,
    labelsep=0.5em,
    itemsep=.5ex,
    ]
    \item\label{dg:events-and-attributes} Display event sequences and heterogeneous tabular attributes in a unified table with one row per item, so events and other data can be compared in one place (\ref{task:select-compare}).
    \item\label{dg:distribution} Show the distribution of the target quantity from similar historical items along with the current item's events (\ref{task:summarize}).
    \item\label{dg:comparison} Enable one-to-many comparison between items for event sequences and other attributes (\ref{task:compare-attributes-similar}, \ref{task:compare-sttributes-candidate}).
    \item\label{dg:align} Enable aligning and sorting by event types for better comparison of all attributes (\ref{task:compare-attributes-similar}, \ref{task:compare-sttributes-candidate}).
\end{enumerate}

\section{Visualization Design}

Over multiple iterations with our collaborators, we designed and implemented a decision-support dashboard that addresses the design goals outlined above.
We first describe the entire dashboard, before providing details on EventColumn.

\subsection{Dashboard Design}

Filters at the top of the dashboard allow filtering by warehouse, shipping date range, and an urgency-flag to exclude coils not suitable for relocation (\ref{task:filter-candidates}).
Below the filters, the main EventColumn visualization (Figure \ref{fig:teaser} top) displays all candidate coils including their events and attributes (\ref{task:compare-sttributes-candidate}) as well as a distribution of the target quantity (\ref{task:summarize}) from \deleted{derived }similar historical items (\ref{task:derive-similar}). \deleted{The similarity was computed in the data preprocessing step based on user-defined attributes such as steel category and coil width.}\inserted{Similar items are passed to EventColumn as lists of IDs per row. For the steel production use case, similar items were precomputed using a proprietary similarity metric based on multiple coil properties.}
When the user selects a coil in the main visualization, another instance of the same visualization at the bottom of the dashboard displays the selected coil's past similar items, allowing direct comparison of their events and attributes, and assessment of the reliability of the storage duration estimate (\ref{task:compare-attributes-similar}).
%TODO: remove step 4?
% After the user has chosen all coils for relocation, an ID table provides their coil IDs for scheduling transport in the planning system (Step~4).

\subsection{EventColumn Design}

Figure \ref{fig:teaser} shows the main visualization including the EventColumn, which is the central contribution of this work.
It extends the Taggle~\cite{furmanova_taggle_2020} tabular visualization technique with a new column type that embeds event sequences directly in the table (\ref{dg:events-and-attributes}).
Each row represents one item. Events are drawn as colored circles in each cell, with color encoding the event type. A shared axis is displayed in the column header.
A boxplot of storage durations for similar historical items (\ref{task:derive-similar}) is overlaid in the same event column (\ref{dg:distribution}).
As all attributes and events share one row per item, they can be compared by visual juxtaposition~\cite{van_der_linden_survey_2023} (\ref{dg:comparison}).
Colors are taken from qualitative ColorBrewer palettes~\cite{brewer_colorbrewer_2003}, using saturated colors for event types and unsaturated colors for categorical columns.

\textbf{Data scale and visual representation.}
As data scale we chose \textit{sequence}~\cite{guo_survey_2022} for comparing all events.
The visual representation is timeline-based and instance-based~\cite{yeshchenko_survey_2024}, using a \textit{fixed} shared time axis as in LifeLines~2~\cite{wang_aligning_2008} and ChronoCorrelator~\cite{van_dortmont_chronocorrelator_2019}. \deleted{Each event type occurs at most once per item.}\inserted{EventColumn expects each event type to occur at most once per item. We discuss this design choice and the resulting limitations in Section~\ref{sec:discussion}.}

\textbf{Overview mode.}
Enabling overview mode (bottom visual in Figure~\ref{fig:teaser}) compresses each row to one pixel in height, allowing hundreds of items to be visible simultaneously (\ref{dg:comparison}).
A selected row remains at full height for detailed inspection, whereas all other rows are condensed.
Compressed event sequences are rendered as small rectangles, and only a single user-chosen boxplot statistic (e.g., median) is shown per row to avoid clutter.

\textbf{Grouping and aggregation.}
Rows can be grouped by any column; grouped event columns are summarized with per-event heatmaps, as shown in Figure~\ref{fig:teaser}-2 (\ref{dg:comparison}).
The time axis is binned and event frequency is encoded by color intensity, inspired by EmoCo's per-channel plots~\cite{zeng_emoco_2020}.
Categorical and numerical columns are aggregated as histograms and boxplots, respectively.

\textbf{Alignment, sorting, and interaction.}
By default, events are placed relative to the current time.
Users can realign the axis to any event type (\ref{dg:align}), similar to LifeLines~2~\cite{wang_aligning_2008} and EventFlow~\cite{monroe_temporal_2013}, making durations between specific events directly comparable across rows. The boxplot can be aligned to any event type, showing the distribution of similar items' storage duration relative to that event.
The event column can be sorted by any event type or boxplot feature, such as shipping time or median storage duration of similar items, to rank items (\ref{dg:align}).
The event column further supports the interactions \emph{scaling} (zoom and pan on the time axis), \emph{segmentation} (show/hide event types), and \emph{emphasis} (column highlighting and per-event color remapping)~\cite{guo_survey_2022,van_der_linden_survey_2023}.

\textbf{Design alternatives considered.}
An earlier design displayed similar coil events in a tooltip on hover rather than in a separate visualization, but was rejected because the second visualization enables more effective user interaction and supports more detailed exploration.
A second draft separated events and tabular data into linked views as in TempViz~\cite{fails_visual_2006}; this was rejected because it breaks the perceptual link between a coil's events and attributes and prevents column-based sorting and grouping (\ref{dg:events-and-attributes}).

\section{Implementation}
The implementation comprises two main parts: extending the Taggle implementation with the EventColumn and integrating the result into a Power BI custom visual. Both parts were implemented in TypeScript.

\subsection{EventColumn}
The \textit{EventColumn} extends Taggle's \textit{MapColumn}, using the event type as map key and a timestamp number as value; boxplot statistics are precomputed and stored under named keys (e.g., \textit{median} and \textit{q1}).
A \textit{getEventValue} method resolves each value relative to a configurable reference event and divides by a configurable time unit (default: one day), so alignment by any event type is handled uniformly for both events and boxplot entries.
A custom cell renderer handles all three Taggle aggregation levels.
At the \textit{summary} level it renders the shared time axis in the column header using a D3~\cite{bostock_d3_2011} linear scale.
At the \textit{group} level each event type is binned by time using D3's binning function, producing a per-event heatmap that encodes frequency by color intensity.
At the \textit{cell} level events are drawn as SVG circles at their scaled horizontal position. 
Boxplots are rendered as SVG lines and rectangles.
In overview mode they are drawn as four-pixel HTML Canvas squares that Taggle can compress to fit more rows.
A sort-dialogue add-on exposes all event types and boxplot statistics as sort keys, and a dedicated event-settings dialogue lets users toggle visible event types and boxplot overlay, and change the reference event.

\subsection{Power BI Custom Visual}
The custom visual is implemented in TypeScript using the Power BI custom visual SDK.
Its data interface defines five input fields: \textit{Data Columns} (general tabular attributes), \textit{Event Data} (one date column per event type), \textit{Similar Data Duration} (semicolon-separated storage durations of similar items), \textit{Similar Data IDs} (IDs of similar items for cross-filtering), and \textit{ID} (a unique row identifier for storing selections).
On each data update from Power~BI, the visual infers the data type of each \textit{Data Columns} field and creates the corresponding Taggle column type (boolean, number, categorical, or date).
\textit{Event Data} columns are converted to milliseconds and passed to the \textit{EventColumn} as event key-value pairs.
For \textit{Similar Data Duration}, the five-number boxplot summary (min, Q1, median, Q3, max) is computed and appended to the row's event data. 
Taggle's state can be saved via a button at the bottom of the visual to preserve user interactions \deleted{such as sorting, grouping, and filtering }across data updates.

The EventColumn Power~BI visual is published at the Microsoft Marketplace~\cite{eventcolumn_marketplace} (under the name EventTableViewer).
The source code and example dashboards are available on GitHub~\cite{eventcolumn_github}. Documentation and usage instructions are available online~\cite{eventcolumn_homepage}.

\section{Use Cases}

In this section, we describe two use cases for EventColumn. The first use case is based on a logistics dataset from our collaborators in the steel industry. For the second use case, we apply EventColumn to the publicly available Brazilian E-Commerce Dataset \cite{olist_andr__sionek_2018}.

\subsection{Steel Production Logistics}

The primary use case motivating this work is the steel production logistics scenario described in Section~\ref{sec:task-abstraction-and-design-goals}: logistics employees must decide which coils to relocate to external storage before the warehouse reaches capacity limits, which requires estimating each coil's remaining storage duration.
The dashboard (Figure~\ref{fig:teaser}) allows users to filter for coils by warehouse and shipping date, and to exclude those with an urgency flag (\ref{task:filter-candidates}).
Then they can compare the displayed coils by their attributes, events (\ref{task:compare-sttributes-candidate}), and storage duration of similar coils (\ref{task:summarize}). If they are unsure about the boxplot distribution's relevance for the current coil, they can select it to see the data of the coils behind the distribution (\ref{task:compare-attributes-similar}). 
Finally, they select coils for relocation based on their experience and the attributes shown in the dashboard.
Because the production data is confidential, Figure~\ref{fig:teaser} uses a synthetic dataset reflecting the main characteristics of the original.

\begin{figure}[tb]
  \centering
  \includegraphics[width=\columnwidth]{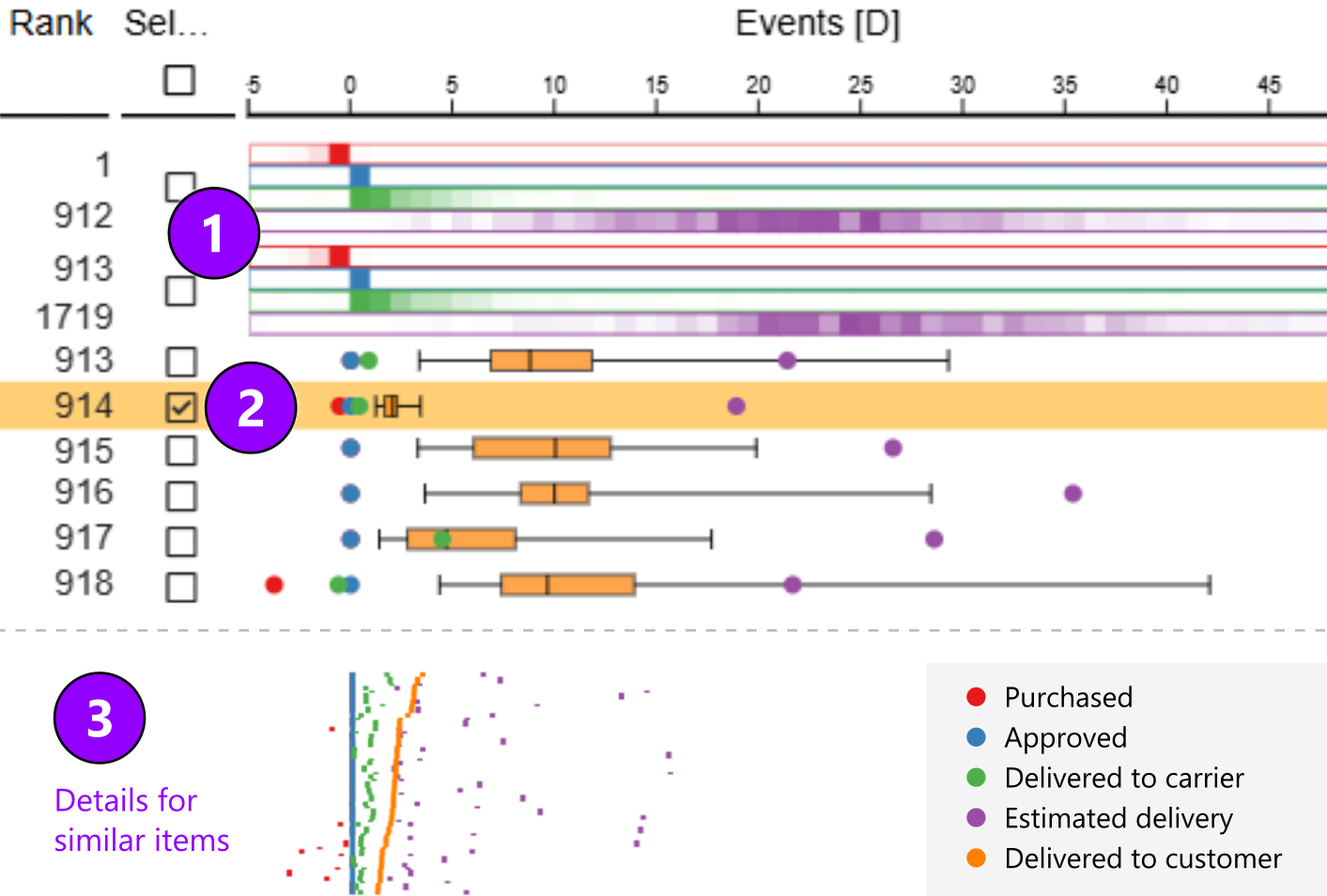}
  \caption{E-commerce analysis dashboard showing delivery events, location, price, and freight value aligned by order approved. We see that the estimated delivery date for items varies between high- and low-value freight~(1). The user selected one order where the estimated delivery date was much later than for recent deliveries to the same city~(2). The analysis of these recent deliveries shows that the estimated delivery date is much later than the actual delivery date~(3).}
  \label{fig:e-commerce-analysis}
\end{figure}

\subsection{Brazilian E-Commerce Dataset}

To demonstrate the EventColumn's capabilities on a publicly available dataset from a different domain, we use the Brazilian E-Commerce Dataset \cite{olist_andr__sionek_2018}, which contains 100,000 orders placed at multiple Brazilian marketplaces between 2016 and 2018.
Each order has the following events: \emph{purchase}, \emph{approved}, \emph{delivered to carrier}, \emph{delivered to customer}, and \emph{estimated delivery date}.
These are accompanied by tabular attributes such as \emph{customer city}, \emph{customer state}, and \emph{freight value}.
The boxplot displays the distribution of the time between \emph{approved} and \emph{delivered to customer} for recent orders to the same city.
Figure~\ref{fig:e-commerce-analysis} shows the resulting dashboard for analyzing delivery times of orders that have not yet been delivered, by adding the duration of the most recent deliveries in the same city to the boxplot data. We first group by freight value to assess whether delivery times differ between high- and low-value freight~(1). This shows significantly lower estimated delivery dates for low-value freight items. We then select one order with a high-value freight where the estimated delivery date is much later than the distribution of recent delivery durations to that city~(2). 
The analysis of recent deliveries to that city\deleted{, sorted by freight value}~(3)\deleted{, also} shows that the estimated delivery date for these orders is later, although all items were delivered within 3 days. This indicates that the selected order will be delivered much earlier, and the estimated delivery date is unreliable for this order.

\section{Discussion}
\label{sec:discussion}

In this section we reflect on preliminary feedback by our collaborators and discuss potential for future work.

\subsection{Preliminary Feedback}

A logistics employee of our collaborator evaluated the dashboard in parallel with the existing tool\inserted{ on real production data} for one month\inserted{---a time range that typically covers decisions for several thousands of coils}. They appreciated \deleted{having an overview of all relevant parameters and events}\inserted{\enquote{all important columns are shown}} in one view. They liked how easy it was to filter for coils in certain width ranges, steel categories, and with a planned galvanizing date. They evaluated their decisions after the coils had been stored and found that they worked well in many cases, but also encountered cases where their estimate was far off. They attributed this to manual planning of the next production steps that does not take the storage location into account. \inserted{In summary, the user concluded that they would like to see \enquote{something similar integrated in a new [logistics] system}, which was in its early planning state at the time of evaluation.}\deleted{The user intends to integrate the dashboard into a newly planned logistics system. They particularly benefited from an overview of similar coils from the past to inform better decision-making.}

% \subsection{Generalization}

% The introduced visualization can be used to efficiently analyze datasets with numeric, categorical, and event sequence attributes with unique event types per item. This is especially useful in production, logistics, or transport, where the process follows an event sequence but is accompanied by information that can influence the events.

\subsection{\inserted{Scalability \& }Limitations\deleted{ and Future Work}}

\inserted{The choice to distinguish event types per color limits EventColumn's applicability to around 10 different types of events, and Power BI's fixed query limit is 1 million rows. Overview mode can help avoid visual clutter in cases of large datasets. However, each item requires at least one pixel of vertical space, and scrolling is introduced if the number of items is larger than the available vertical pixel space. The width of each event mark in overview mode is constrained to four pixels, which can lead to problems when events are very close together within a very large overall time range.}

\deleted{EventColumn can be used to analyze datasets with numeric, categorical, and event sequence attributes with unique event types per item.
However, the main limitation}\inserted{Currently, the main conceptual limitation of EventColumn} is that recurring events per item are not supported. \inserted{Enabling alignment and sorting of rows would require more complex interactions or aggregation. As a workaround, multiple events can be aggregated in preprocessing or stored with different labels in cases of fixed event counts per item.}\deleted{Implementing this poses the challenge of aligning and sorting by event type when event counts differ across items.}
\deleted{Another limitation is filtering by events, which is currently only possible when adding the event as a separate date column. Also, the boxplot currently only supports loading the duration distribution between a single fixed source event and a single target event.
These limitations are not conceptual, but arise from details of the underlying Taggle system~\cite{furmanova_taggle_2020}, which we used as a basis for the current implementation.}\inserted{Other minor limitations---e.g., that filtering requires adding an event as a separate date column---result directly from basing EventColumn on the Taggle system~\cite{furmanova_taggle_2020}.}

\section{Conclusion}

This paper introduced the EventColumn: a new column type for tabular visualizations that enables analysis of event sequences, numerical, and categorical data without switching between visualizations. Event sequences can be compared both at the instance and group level. Aligning and sorting by event type, in combination with duration distributions from past events, enables informed decision-making. We demonstrated the visualization's capabilities on datasets from two domains. Our work shows that general-purpose tabular frameworks can be extended to natively accommodate event sequence data\deleted{ while preserving their sorting, grouping, and comparison capabilities---suggesting}\inserted{, illustrating} a path toward richer mixed-type column libraries for interactive data exploration.
%The visualization can be enhanced to support recurring events, advanced filtering, and multi-encoding of event type in the future.

\acknowledgments{
This work has been supported by the BMIMI, BMWET, and FFG, Contract No. 911655: \enquote{Pro²Future II}.}

\bibliographystyle{abbrv-doi-hyperref}

\bibliography{references}
\end{document}